\documentclass[aps,prb,twocolumn,groupedaddress,showpacs]{revtex4}

\usepackage{graphicx}

\begin{document}


\title{Time-resolved impulse response of the magnetoplasmon resonance in a two-dimensional electron gas}


\author{E.A. Shaner}
\author{S.A. Lyon}
\affiliation{Princeton University Electrical Engineering Department}


\date{\today}

\begin{abstract}
We have used optically excited ultrashort electrical pulses to measure the magnetoplasmon resonance of a two-dimensional electron gas formed in an AlGaAs/GaAs heterostructure at frequencies up to 200 gigahertz.  This is accomplished by incorporating the sample into a guided wave probe operating in a pumped \( ^{3}He \) system.  We are able to detect the resonance by launching a stimulus pulse in the guide, and monitoring the system response in a time resolved pump-probe arrangement.  Data obtained from measurements yield resonant frequencies that agree with the magnetoplasmon dispersion relation.
\end{abstract}


\maketitle

\section{INTRODUCTION}
The far infrared properties of two-dimensional electron gas (2DEG) systems have been studied extensively.  Common examples are observations of cyclotron resonance\cite{ref1,ref2}, magnetoplasmon resonance\cite{ref3,ref4}, and edge magnetoplasmons\cite{ref5}.  Here we address a new approach to these measurements and demonstrate the time resolved detection of the magnetoplasmon resonance.  Normally, it is necessary to fabricate a grating on the top of the sample in order to allow coupling of far infrared radiation to the plasmon component of a magnetoplasmon mode\cite{ref5}.  One then excites with continuous wave radiation and looks at a particular optical property, such as transmission, which averages the time dependent behavior of the carriers.  With our technique, the grating restriction is removed and we are able to directly time resolve the effect of the applied field on the resonance.  
	
	The magnetoplasmon resonance is formed by the coupling of the two-dimensional plasmon and the cyclotron motion of carriers.  The cyclotron resonance frequency is given by\cite{ref6} \( \omega _{c} =eB/m^{*} \) , where \(e\) is the electron charge, \(B\) is the perpendicular component of the magnetic field, and \(m^{*}=.067m_{e}\) is the effective mass of an electron in GaAs.  The dispersion relation for 2D plasmons is given approximately by\cite{ref7} \( \omega _{p}^{2} \)=\( e^{2}n_{2D}q/(2\varepsilon _{0}\varepsilon _{b}m^{*}) \) where \(n_{2D}\) is the two dimensional density of electrons, q is the 2D plasmon wave vector, and \( \varepsilon _{b} \) is the dielectric constant of GaAs. When a magnetic field is applied to a 2DEG, the cyclotron and plasmon couple yielding a magnetoplasmon mode with a resonant frequency\cite{ref5} \(\omega ^{2}=\omega _{p}^{2}+\omega _{c}^{2}\).

	A novel method for measuring conductivity resonances in 2DEG systems is to incorporate the 2DEG as part of a three conductor coplanar waveguide (CPW)\cite{ref8}.  The fringing fields from the conductors capacitively couple to the 2DEG allowing it to affect waveguide losses and dispersion.  Typically, to perform measurements at frequencies up to 30GHz, one feeds microwaves to the CPW through a coaxial cable, and then either measures transmitted power with a detector near the sample\cite{ref8}, or couples the microwaves back out of the system for analysis of amplitude and phase shift\cite{ref9}.  The guiding of microwaves in and out of cryogenic systems, as well as impedance matching to the CPW on the sample in order to eliminate back reflections, can be an arduous task.  With this approach, one is also limited to frequencies that can be generated through microwave electronics.  In order to circumvent these difficulties we used a similar waveguide, two-conductor coplanar strip transmission line (CPS), excited optoelectronically through illumination of Auston switches\cite{ref10}.  In this work, Auston switches are formed by patterning conductors on low temperature grown GaAs (LT GaAs).  This material has recombination time of approximately one picosecond\cite{ref11} and allows for the generation and detection of electrical pulses on that timescale.  In contrast to other experiments which utilized optoelectronic excitation\cite{VonKlitz}, we also implement optoelectronic sampling of the waveguide signal using a second Auston switch integrated on the waveguide. 
\section{EXPERIMENT}	
	The CPS consisted of two 6\( \mu m \) wide, 50$\AA$/2000$\AA$ thick Ti/Au strips separated by 10\( \mu m \).  These strips were patterned on a multilayer substrate consisting of a 2\( \mu m \) LT GaAs epilayer, grown by molecular beam epitaxy (MBE), which has been transferred onto a 50\( \mu m \) thick polyimide sheet using the epoxy bond and stop etch process\cite{ref12}.  The entire structure is then mounted on a silicon substrate which serves as structural support and also allows for the attachment of optical fibers.  An MBE grown AlGaAs/GaAs 2DEG with mobility at 4K of \(6\cdot 10^{5}cm^{2}V^{-1}s^{-1}\) and density \(n_{2D}=2.8\cdot 10^{11}cm^{-2}\) was employed as a superstrate.  Figure 1 shows the resulting waveguide and sample structure.  The electric field mode for this type of guide is a dipole pattern\cite{ref13}.  The fringing fields of this dipole allow for capacitive coupling to the 2DEG.  Single mode optical fibers are used to carry 780nm wavelength pump and probe pulses of 3ps duration generated by a Ti:Sapphire laser to the CPS.  To launch a pulse, a bias is placed across the CPS conductors, as shown in Fig. 1c.  When the pump pulse strikes the area between the CPS conductors, charge displacement and collection, driven by the the DC bias, efficiently excites the CPS mode\cite{ref13}.  Due to the short carrier lifetime of the LT GaAs, this produces an electrical signal with approximately the shape of the optical pulse envelope\cite{ref10}.  The voltage on the CPS is sampled in time through a similar process by striking the sampling gap, shown in Fig. 1c, with a probe pulse.  By delaying the probe pulse with respect to the pump, and monitoring the sampling line signal, we time resolve pulse propagation\cite{ref14}.  This gives us a measurement of the impulse response of the entire waveguiding structure.
 \begin{figure}

 \resizebox*{8.6cm}{!}{\includegraphics{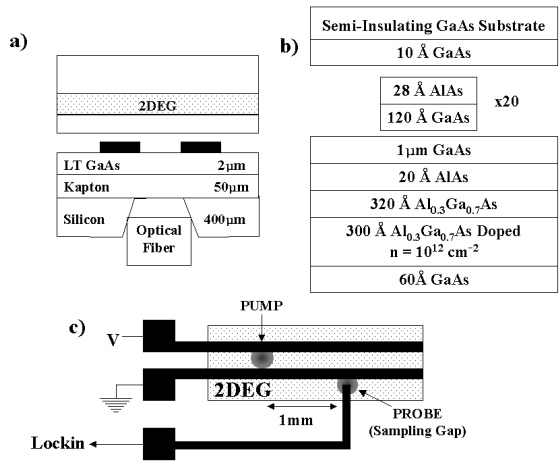}}
 \caption{\label{Experimental Setup}(a) The 2DEG sample applied as a waveguide superstrate.  The growth structure of the 2DEG used in this experiment is shown in (b).  Optical pulses are delivered to the waveguide sample through optical fibers in order to perform pump-probe measurements.  This is shown in (c) along with the electrical connections.}
 \end{figure}

	Due to dispersion in the optical fiber, the pulses are broadened by roughly 1ps.  Measurements shown in Fig. 2 were taken with approximately 300\( \mu W \) total laser power in a pumped \( ^{3}He \) system at T=0.5K.  The CPS was biased at 5V giving reasonable signal-to-noise ratio.  The pump beam was modulated at 1KHz to allow for lockin detection of the sampling line signal.  As shown, the data can be broken up into three regions.  The section labeled initial pulse is the first pass of the pulse launched by the pump beam after it has traveled 1mm from the pump gap to the sampling gap.  Trailing the initial pulse is a distinctive ringing due to the magnetoplasmon resonance\cite{ref15}.  Since our transmission lines did not have an impedance matched termination, there is a large reflection when the pulse reaches the end of the CPS.  The location of this reflection is indicated in the data, although most of it has been truncated since it does not contain easily interpreted information.  This reflection limits the time window in which we can observe the system response to 30ps in the present structures.
\begin{figure}
\resizebox*{8.6cm}{!}{\includegraphics{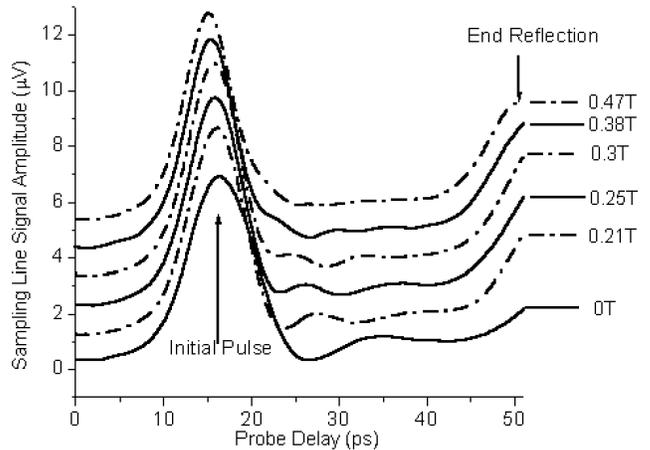}}
\caption{Sampling line signal as measured through a lockin amplifier.  Using a standard optical delay line, the Probe pulse is delayed relative to Pump in order to provide time resolution of measurement.\label{scans}}
\end{figure}
\section{ANALYSIS}
\begin{figure}
\resizebox*{8.6cm}{!}{\includegraphics{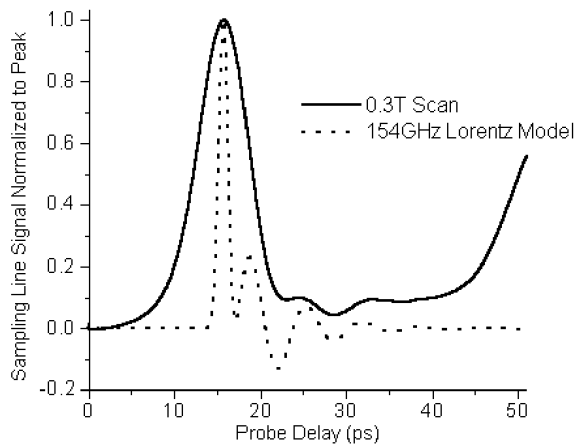}}
\caption{Simulation of pulse propagation in a medium containing Lorentz oscillators along with a real scan taken a B=0.3T.  Simualted pulse was chosen to be 0.5ps FWHM in order to show additional ringing.  \label{Lorentz}}
\end{figure}
	To gain an approximate understanding of how pulses should propagate with the 2DEG applied as a superstrate, we modeled our system with a dielectric constant of a medium containing Lorentz oscillators.  For simplicity, we neglected modal dispersion and used a quasi-TEM approximation\cite{ref15}.  A short simulated pulse propagated in this manner is shown in Fig. 3 along with data taken at B=0.3T.  The frequency of the modeled oscillator was chosen to be 154GHz to match the oscillations of the data in the response region.  From the calculations, the first peak we see after the initial pulse should appear at approximately 1.5 periods of the resonant frequency.  This is important information since our pulses are broad enough to hide the first ring of the resonance.  With this knowledge, peaks found in the the time domain data are converted into resonant frequencies.  The data shown in Fig. 2 at zero magnetic field shows that we are exciting a 2D plasmon of approximately 80GHz.  A simple estimate using \(q=2\pi /\lambda \) and taking \( \lambda =20\mu m \) (corresponding to a half wavelength of the plasmon being the 10\( \mu m \) line spacing) gives 200GHz. An exact agreement is not expected as the CPS conductors and DC bias likely modify the 2D plasmon dispersion\cite{ref7}.  

	Frequencies obtained from the time domain data of Fig. 2, along with other scans not shown, are plotted against perpendicular magnetic field in Fig. 4.  The resulting dispersion shows that our measurements follow the expected magnetoplasmon curve.  This verifies that our technique can provide information about dipolar resonances without the need of contacting the sample.  Although the resonant frequency is readily attained from our data, accurately quantifying the damping of the resonance would require a longer time window.

	In the data taken at B=0.38T, it is possible to resolve the decay of the resonance and determine a relaxation time.  In the time domain, we fit an exponential decay to the rings, and in the frequency domain, we fit a lorentian lineshape to the power spectrum.  Both methods produced a relaxation time of approximately 5ps which is signifcantly lower than the 22ps scattering time obtained from the sample mobility.  This result is expected since the cyclotron mobility at low fields is more sensitive to small angle scattering than is the transport mobility\cite{DasSarma}.

\begin{figure}
\resizebox*{8.6cm}{!}{\includegraphics{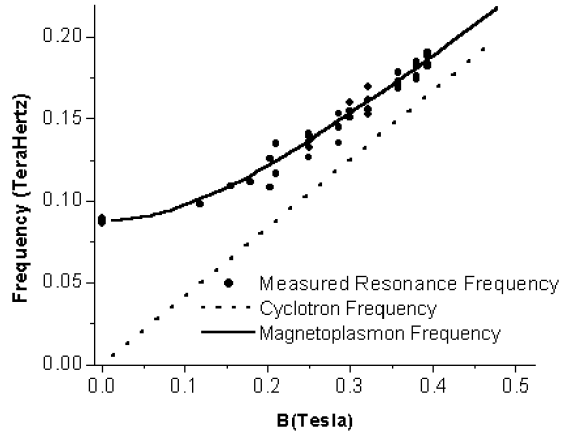}}
\caption{Dispersion as obtained from pulse measurements compared to theoretical magnetoplasmon dispersion.  Cyclotron dispersion is added as a reference.\label{freq}}
\end{figure}

\section{CONCLUSION}
	With our time-resolved measurement system, we were able to characterize 2DEG resonances as a function of magnetic field.  Resonant frequencies obtained from the time domain data presented here match the expected magnetoplasmon dispersion curve well.  As discussed, these measurements were taken with 300\( \mu W \) total laser power resulting in a minimum measurement temperature of T=0.5K in our system.  Measurements have been taken with as little as 70\( \mu W \) total laser power yielding lower temperatures.  In the present experiment, temperature was not critical and higher power gave better signal-to-noise.  Although the data shown was taken at magnetic fields below 1T, we note that the probe itself had no problems operating at fields up to 8.5T, which was limited only by the magnet in our system.  Future design improvements will provide a larger time window in which more precise characterization of the resonance ring-down will be possible.

\begin{acknowledgments}
We would like to thank Mike Valenti of Princeton University for assistance in preparation of silicon substrates used in these experiments.  E.A. Shaner was supported by a National Defense Science and Engineering graduate fellowship.  This work was supported in part by the U.S. Army Research Office under grant \(\#\)DAAG55-98-1-0270.
\end{acknowledgments}
\bibliography{paper}
\end{document}